\documentclass[11pt]{article}

\pdfoutput=1

\usepackage[makeroom]{cancel}
\usepackage{color}
\usepackage{graphicx}
\usepackage{amsmath}
\usepackage{amssymb}
\usepackage{xspace}
\usepackage[small]{subfigure}
\usepackage[numbers,compress]{natbib}
\usepackage[hyperfootnotes=false]{hyperref}
\usepackage{tcolorbox}

\newlength{\fighskip} \fighskip=2pt
\newlength{\figvskip} \figvskip=3pt

\newcommand*{\figbox}[2]{{
  \def\figscale{#1}
  \def\arraystretch{0.8}
  \arraycolsep=0pt
  \begin{array}{c}
    \vbox{\vskip\figscale\figvskip
      \hbox{\hskip\figscale\fighskip
        \includegraphics[scale=\figscale]{#2}}}
  \end{array}}}

\DeclareMathOperator{\Tr}{Tr}

\usepackage{mciteplus} 
\usepackage{dcolumn}
\usepackage{bm}
\usepackage{verbatim}
\usepackage{amscd}
\usepackage{amsfonts}
\usepackage{setspace}
\usepackage{amsthm}
\usepackage{enumerate}
\usepackage{mathtools}

\theoremstyle{plain}
\newtheorem{theorem}{Theorem}
\theoremstyle{plain}
\newtheorem{lemma}{Lemma}
\theoremstyle{plain}
 
\theoremstyle{plain}

\theoremstyle{remark}

\begin{document}

\title{\bf 
Recovery algorithms for Clifford Hayden-Preskill problem 
}
\author{
Beni Yoshida\\ 
{\em \small Perimeter Institute for Theoretical Physics, Waterloo, Ontario N2L 2Y5, Canada} }
\date{}

\maketitle

\begin{abstract}
The Hayden-Preskill recovery problem has provided useful insights on physics of quantum black holes as well as dynamics in quantum many-body systems from the viewpoint of quantum error-correcting codes. 
While finding an efficient universal information recovery procedure seems challenging, some interesting classes of dynamical systems may admit efficient recovery algorithms.
Here we present simple deterministic recovery algorithms for the Hayden-Preskill problem when its unitary dynamics is given by a Clifford operator.
The algorithms utilize generalized Bell measurements and apply feedback operations based on the measurement result. 
The recovery fidelity and the necessary feedback operation can be found by analyzing the operator growth.  
These algorithms can also serve as a decoding strategy for entanglement-assisted quantum error-correcting codes (EAQECCs).
We also present a version of recovery algorithms with local Pauli basis measurements, which can be viewed as a many-body generalization of quantum teleportation with fault-tolerance.
A certain relation between out-of-time order correlation functions and discrete Wigner functions is also discussed, which may be of independent interest.
\end{abstract}

\vspace{-0.7\baselineskip}

\section{Introduction}

The Hayden-Preskill thought experiment addresses a fundamental question concerning whether quantum information can escape from a quantum black hole~\cite{Hayden07, Hayden:2008aa}. 
Recent progresses concerning the Hayden-Preskill thought experiment have led to a number of interesting results in various branches of theoretical physics. 
In studies of quantum gravity, it has provided useful insights on the black hole information loss puzzle~\cite{Gao:2017aa, Traversable2017, Penington19, Almheiri19}, the entanglement structure of the black hole interior~\cite{Verlinde12, Yoshida:2021aa, Beni18, Beni19, Yoshida:2020aa, Pasterski20} and the quantum error-correcting nature of holography~\cite{Almheiri:2015ac, Pastawski15b, Dong2016, Cotler:2019aa, Hayden:2016aa}, and has led to solvable and tractable toy models of the AdS/CFT correspondence~\cite{Kitaev_unpublished, Maldacena:2016ab, Maldacena:2016aa}. 
In studies of condensed matter physics, the notion of quantum information scrambling has also improved our understanding of dynamical nature of quantum entanglement in many-body quantum systems~\cite{Hartman13, Hosur:2015ylk, Nahum:2017aa, Nahum:2018aa, Khemani:2018aa, Rakovszky:2018aa, Keyserlingk:2018aa, Li:2018aa, Skinner:2019aa}. 
These developments have further stimulated researches on pseudo-randomness and quantum error-correcting codes in quantum information theory~\cite{Harrow:2009aa, Brandao:2016aa, Brown:2015aa, Cotler:2017aa, Brandao19, hunterjones2019unitary, Roberts:2017aa, Nakata:aa, Bao:2021aa}.

While the original work by Hayden and Preskill was restricted to Haar random unitary operators (or unitary operators drawn from an ensemble forming an approximate unitary $2$-design), subsequent works showed that a much larger class of unitary dynamics $U$ achieves the recovery phenomena. Namely, we proved that, if the unitary dynamics $U$ is scrambling, then the Hayden-Preskill recovery is possible~\cite{Hosur:2015ylk}. Here the notion of scrambling can be rigorously defined by utilizing out-of-time ordered correlation (OTOC) functions of the form $\langle O_{A}(0)O_{D}(t)O_{A}(0)O_{D}(t) \rangle$, see~\cite{Yoshida:2017aa} for details. In the context of the Hayden-Preskill recovery problem, $A$ and $D$ correspond to the input degrees of freedom and the late Hawking radiation respectively. Since the dynamics of quantum black holes is scrambling in a sense of the OTOC decay, this result provides a formal information theoretic proof that a quantum information indeed escapes from an old quantum black hole. It is worth emphasizing that scrambling is not a necessary condition for the recoverability in the Hayden-Preskill problem as we shall discuss later.

Another important question concerns constructions of the recovery channel $\mathcal{R}$ in the Hayden-Preskill problem. An explicit probabilistic recovery algorithm, which works universally for scrambling systems, has been proposed by the author and Kitaev~\cite{Yoshida:2017aa}. While the success probability of this algorithm scales as $O\big(\frac{1}{d_{A}^2}\big)$ with $d_{A}$ being the Hilbert space dimension of the input state, one can convert this probabilistic algorithm to a deterministic one by the use of the Grover search algorithm with the quantum circuit complexity increasing by a factor of $O(d_{A})$. Hence, this generic algorithm does not work efficiently when the input Hilbert space size scales exponentially with respect to the system size $n$. To the best of our knowledge, no efficient recovery algorithm is currently known for such cases. Other recovery algorithms, which work for more specialized systems (such as the SYK model only in the early time/low temperature regime before the scrambling time), have been also proposed~\cite{Brown:aa, Nezami:aa, Gao:aa, Schuster:aa}. These algorithms work deterministically without the use of the Grover search routine, and are inspired by the traversable wormhole geometries in the two-sided AdS black hole. It is however worth noting that, in these traversable wormhole protocols, the number of qubits to be collected from the late Hawking radiation ($D$) needs to be much larger than the input qubits ($A$) and often scales with the system size $n$, as opposed to the original findings by Hayden and Preskill which only require $D$ to be slightly larger than $A$.

The Hayden-Preskill recovery problem can be understood as an example of entanglement-assisted quantum error-correcting codes (EAQECCs) ~\cite{Bennett:1999aa, Brun:2006ab}. In EAQECCs, a sender and a receiver share quantum entanglement a priori where qubits on a receiver's end are assumed to be free from errors. The encoded information tends to be robust against various strong forms of errors acting on the qubits on a sender's end, and achieves various advantages over conventional quantum error-correcting codes. In the Hayden-Preskill problem, the initial EPR pairs between an old black hole $B$ and the early radiation $\overline{B}$ can be identified as the pre-shared entanglement where $\overline{B}$ corresponds to the receiver's share. The erasure error acts on qubits on $C$ (the remaining black hole) on the sender's end while we send qubits on $D$ (the late radiation) to the receiver. Note that the framework of EAQECCs does not require the entanglement resource $B\overline{B}$ to be maximally entangled. As such, the Hayden-Preskill recovery problem should be understood as a subclass of EAQECCs. Since the Hayden-Preskill recovery problem can be viewed as the stabilizer-based entanglement-assisted quantum error-correcting codes (often abbreviated as EAQECCs)~\cite{Bennett:1999aa, Brun:2006aa}, solving the Clifford Hayden-Preskill recovery problem will provide an efficient decoding method for stabilizer-based EAQECCs.

\subsection{Main results}

In this paper, we study the Hayden-Preskill recovery problem when its unitary time-evolution is supplied by a Clifford operator. While analyzing dynamical properties of interacting quantum many-body systems is a notoriously challenging problem in general, the Clifford dynamics serves as analytically and computationally tractable toy models of quantum many-body dynamics. Hence, finding an explicit recovery algorithm for the Clifford Hayden-Preskill problem will provide further insights onto scrambling and thermalization phenomenon arising in quantum many-body systems, and may reveal an important hint toward the verification of dynamically generated entanglement. 

Namely, we will present simple and efficient deterministic recovery algorithms for the Clifford Hayden-Preskill problem. The recovery algorithm runs with some modification to the one from~\cite{Yoshida:2017aa} where (generalized) Bell measurements are performed on pairs of the late Hawking radiation. By applying an appropriate feedback Pauli operators on the output qubits, one can deterministically reconstruct the initial input state without the need of implementing the Grover search algorithm. The feedback Pauli operators can be found by computing the operator growth, which can be performed efficiently on a classical computer. We also find explicit expressions of logical operators by studying how a local Pauli operator time-evolves backwards to the past. Note that it is well known that a stabilizer code under the erasure channel can be decoded efficiently by using the standard Gottesman-Knill treatment with Gaussian elimination. Unlike the standard treatment, our recovery strategy is concrete and physically motivated. 

In order to obtain the aforementioned results, we employ a certain relation between discrete Wigner functions and OTOC functions. This relation may be of independent interest from resource theoretic viewpoints. We speculate that OTOC functions may serve as a useful probe of non-Cliffordness. 

We will also provide a version of the recovery algorithms based on local Pauli measurements, instead of measurements with entangled basis states. In this algorithm, the sender and receiver perform local measurements on their shares of qubits, and only the classical communication is transmitted from the sender to the receiver. This recovery algorithm can be viewed as a novel many-body generalization of quantum teleportation which utilizes quantum error-correcting codes~\footnote{Protocols from~\cite{Brown:aa, Nezami:aa, Gao:aa, Schuster:aa} also achieve many-body versions of quantum teleportation, but the decoding strategies are different.}. Fault-tolerance of the scheme may be desirable in performing quantum teleportation in a noisy environment. It is also worth mentioning that the Hayden-Preskill recovery problem with local basis measurements is fundamentally akin to how the volume-law entanglement is dynamically generated in monitored quantum circuits as discussed in detail in an accompanying work~\cite{BYoshida21a}. 

This paper is organized as follows. In section~\ref{sec:review}, we present a brief review of the Hayden-Preskill problem. In section~\ref{sec:Clifford}, we derive a necessary and sufficient condition for the recoverability in the Clifford Hayden-Preskill problem. In section~\ref{sec:recovery}, we present a recovery algorithm with generalized Bell measurement. In section~\ref{sec:logical}, we construct logical operators. In section~\ref{sec:local}, we present a version of a recovery algorithm with local Pauli measurement. In section~\ref{sec:outlook}, we conclude with discussions. 

\section{Hayden-Preskill problem}\label{sec:review}

In this section, we present a brief review of the Hayden-Preskill problem~\cite{Hayden07}. While the problem originally stems from consideration of quantum aspects of black hole physics, it can be rigorously formulated by utilizing the quantum information theoretic language. Throughout the paper, we will assume that the system consists of qubits since generalizations to systems with qudits (multi-state spins) are straightforward. 

\subsection{Statement of the problem}

Assume that an unknown input quantum state $|\psi\rangle_{A}$ is prepared on a Hilbert space $A$ which consists of $n_{A}$ qubits. In addition, $n_{B}$ copies of EPR pairs are prepared on the Hilbert space $B$ and $\overline{B}$. Explicitly, the EPR pairs on $B\overline{B}$ are given by 
\begin{align}
|\text{EPR}\rangle_{B\overline{B}} = \frac{1}{\sqrt{d_{B}}} \sum_{j=1}^{d_{B}} |j\rangle_{B} \otimes |j\rangle_{\overline{B}},\qquad d_{B} = 2^{n_{B}}
\end{align}
where each of $B,\overline{B}$ consists of $n_{B}$ qubits. Hence, our initial state is given by
\begin{align}
|\psi\rangle_{A} \otimes |\text{EPR}\rangle_{B\overline{B}}.
\end{align}
In general, the entangled state on $B\overline{B}$ can be thermofield double states of the underlying Hamiltonian $H$, but we will use $|\text{EPR}\rangle$ as we will treat the Clifford evolution which is not associated with a Hamiltonian.

The system undergoes the time evolution by a unitary operator $U$ which acts non-trivially only on $AB$. The outcome of the time evolution is 
\begin{align}
U_{AB}\otimes I_{\overline{B}} \big(|\psi\rangle_{A} \otimes |\text{EPR}\rangle_{B\overline{B}} \big)
\end{align}
where the time evolution acts trivially on $\overline{B}$. We think of splitting the Hilbert space $\mathcal{H}=\mathcal{H}_{A}\otimes \mathcal{H}_{B}$ into two subsystems $CD$:
\begin{align}
\mathcal{H}=\mathcal{H}_{A}\otimes \mathcal{H}_{B} = \mathcal{H}_{C}\otimes \mathcal{H}_{D}.
\end{align}
Here $A,B,C,D$ consist of $n_{A},n_{B},n_{C},n_{D}$ qubits with $n=n_{A}+n_{B} = n_{C}+n_{D}$, and $\mathcal{H}_{A}\otimes \mathcal{H}_{B} = \mathcal{H}_{C} \otimes \mathcal{H}_{D}$ represent the same Hilbert space with different partitionings into $A,B$ and $C,D$ respectively. In these partitionings, the unitary time evolution operator acts as follows:
\begin{align}
U : \mathcal{H}_{A}\otimes \mathcal{H}_{B} \rightarrow \mathcal{H}_{C} \otimes \mathcal{H}_{D}
\end{align}
The outcome of the time evolution can be graphically shown as follows:
\begin{align}
\figbox{1.2}{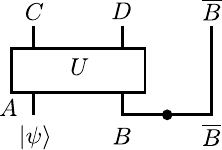}\ .
\end{align}
Here we employed a tensor diagram to represent the wavefunction graphically where the time evolves upward. A horizontal line on $B$ and $\overline{B}$ represents the EPR pairs and a black dot in the diagram represents a factor of $1/\sqrt{d_{B}}$ in $|\text{EPR}\rangle_{B\overline{B}}$ for proper normalization.

The Hayden-Preskill recovery problem asks whether one can reconstruct the input state $|\psi \rangle_{A}$ by having an access to $D$ and $\overline{B}$. The above time-evolution can be interpreted as an encoding of a quantum error-correcting code with an isometry $\Gamma$:
\begin{align}
\Gamma(|\psi\rangle_{A}) \equiv U_{AB}\otimes I_{\overline{B}} \big(|\psi\rangle_{A} \otimes |\text{EPR}\rangle_{B\overline{B}} \big) \qquad \Gamma: \mathcal{H}_{A}\rightarrow \mathcal{H}_{C}\otimes \mathcal{H}_{D} \otimes \mathcal{H}_{\overline{B}}.
\end{align}
The Hayden-Preskill recovery problem is then identical to the decoding problem of a quantum error-correcting code that undergoes an erasure channel which removes qubits on $C$.  The total (noisy) quantum channel $\mathcal{N}$ can be defined as
\begin{align}
\mathcal{N}(\sigma_{A}) = \Tr_{C}\big( \Gamma (\sigma_{A}) \big) \qquad \mathcal{N}: \mathcal{H}_{A}\rightarrow \mathcal{H}_{D} \otimes \mathcal{H}_{\overline{B}}.
\end{align}
Our task is to understand whether a recovery channel $\mathcal{R}$ with $\mathcal{R}(\mathcal{N}(\sigma_{A}))\approx \sigma_{A}$ exists or not, and to construct an explicit recovery channel $\mathcal{R}$ if exists~\footnote{In a regime where the Hayden-Preskill recovery is possible, the Petz recovery map can be employed as a recovery map since the relative entropy difference is small. In~\cite{in_prep}, it was pointed out that the Petz map can be reduced to a simpler recovery algorithm from~\cite{Yoshida:2017aa} due to the decoupling phenomena.}.  

In the context of black hole physics, the initial state $|\psi\rangle_{A}$ is interpreted as an object which falls into a black hole. In the HP thought experiment, instead of a newly formed black hole, an old black hole, which has already emitted more than half of its content via the Hawking radiation, is considered. An old black hole $B$ is assumed to be maximally entangled with the early radiation $\overline{B}$, so it is typically modelled as multiple copies of EPR pairs $|\text{EPR}\rangle_{B\overline{B}}$.
Here the number of qubits $n_{B}$ corresponds to the Bekenstein-Hawking entropy of the black hole. Obviously, the actual old black hole will not look like $n_{B}$ copies of EPR pairs. Here we think of distilling $n_{B}$ qubits which are maximally entangled with the early radiation $\overline{B}$ and model the distilled quantum state as EPR pairs. In this sense, $n_{B}$ should be thought of as the coarse-grained entropy of the black hole.

The black hole dynamics $U$ mixes the input state $A$ and the old black hole $B$ unitarily. Later, the Hawking radiation $D$ is emitted from the black hole while $C$ corresponds to the remaining black hole. The Hayden-Preskill recovery problem asks whether the initial state $|\psi\rangle_A$ can be reconstructed by accessing the early and late Hawking radiations $\overline{B}$ and $D$ while the remaining black hole $C$ is still inaccessible. 

\subsection{Decoupling}

To discuss the recoverability of the initial state, it is convenient to purify the input state by appending the reference system $R$ with $|\text{EPR}\rangle_{RA}$.  Namely, we will consider the following quantum state:
\begin{align}
|\Psi\rangle_{RCD\overline{B}} &= (I_{R} \otimes U_{AB} \otimes I_{\overline{B}}) |\text{EPR}\rangle_{RA} \otimes |\text{EPR}\rangle_{B\overline{B}} 
= \ \figbox{1.2}{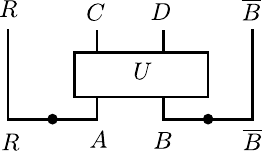}\ . \label{eq:state}
\end{align}
Note that projecting $R$ onto $|\psi^{*}\rangle_{R}$ prepares $|\psi\rangle_{A}$ on $A$, and reduces the system to the situation originally considered in the previous subsections.

Recall that the mutual information for a pure quantum state obeys the following relation:
\begin{align}
I(R,C) + I(R,D\overline{B}) = 2 S_{R} = 2 \log d_{R}. \label{eq:duality}
\end{align}
A perfect recovery is possible if and only if $R$ is maximally entangled with $D\overline{B}$~\footnote{Here, by perfect recovery, we mean that there must exist a perfect recovery channel $\mathcal{R}$ which works for an arbitrary input state $|\psi\rangle_{A}$.}.  In this case, due to Eq.~\eqref{eq:duality}, we have $I(R,C)=0$, so the reference $R$ will be decoupled from $C$:
\begin{align}
\rho_{RC} = \rho_{R}\otimes \rho_{C} \qquad \rho_{R} = \frac{I_{R}}{d_{A}} \quad \rho_{C} = \frac{I_{C}}{d_{C}}
\end{align}
where $\rho_{RC}$ is defined with respect to the purified state in Eq.~\eqref{eq:state}. Hence, we have the following criteria.

\begin{lemma}\label{lemma:criteria}
A perfect recovery in the Hayden-Preskill problem is possible if and only if 
\begin{align}
S_{RC} = n_{A} + n_{C}.
\end{align}
\end{lemma}

A Clifford operator is a unitary operator $U$ which satisfies the following property:
\begin{align}
U P U^{\dagger} \in \mathrm{Pauli} \qquad \text{where $P \in \mathrm{Pauli}$}.
\end{align}
In other words, it transforms Pauli operators into Pauli operators up to a phase factor. When the dynamics of the Hayden-Preskill problem is given by a Clifford operator, the resulting quantum code, associated with the isometry $\Gamma$, is a stabilizer code. Stabilizer generators of the code can be found by noticing that the following Pauli operators stabilize $|\text{EPR}\rangle_{B\overline{B}}$:
\begin{align}
X_{B_{i}}\otimes X_{\overline{B_{i}}}, \qquad Z_{B_{i}}\otimes Z_{\overline{B_{i}}} \qquad i = 1,\cdots, n_{B}.
\end{align}
Hence, stabilizer generators are given by 
\begin{align}
U(X_{B_{i}}\otimes X_{\overline{B_{i}}})U^{\dagger}, \qquad U(Z_{B_{i}}\otimes Z_{\overline{B_{i}}})U^{\dagger}.
\end{align}

We will be particularly interested in finding expressions of logical operators. One possible expression of a logical operator can be found by time-evolving Pauli operators supported on $A$ by $U$:
\begin{align}
\widetilde{P_{A}}=UP_{A}U^{\dagger}.
\end{align}
By construction, this expression of a logical operator $\overline{P_{A}}$ is supported on $CD$. Recall however that expressions of a logical operator are not unique, and are equivalent up to applications of stabilizer operators. In the framework of stabilizer codes, the duality relation in Eq.~\eqref{eq:duality} can be understood as the so-called cleaning lemma~\cite{Bravyi09, Haah10, Beni10}, see~\cite{Li:2021aa} also. Namely, when $R$ and $C$ are decoupled, no logical operator can be supported on $C$. This in turn implies that all the logical operators can be supported on $D\overline{B}$ (in other words, supports on $C$ can be always ``cleaned'' so that logical operators are supported exclusively on $D\overline{B}$). In section~\ref{sec:logical}, we will present explicit constructions of logical operators supported on $D\overline{B}$.

\section{Recoverability}\label{sec:Clifford}

In this section, we derive the necessary and sufficient condition for perfect recovery for the Clifford Hayden-Preskill problem by looking at how local Pauli operators $P_{A}$ on $A$ time-evolves and overlap with $D$. Namely, we prove that perfect decoupling $\rho_{RC}=\rho_{R}\otimes \rho_{C}$ occurs if and only if each of $P_{A}$ evolves into $P_{A}(t)$ with unique Pauli operator profile on $D$.

\subsection{Forward time-evolution map}

Recall that there are $d_{A}^2$ different Pauli operators on $A$. For each Pauli operator $P_{A}$, let us consider its time-evolution $P_{A}(t) = U P_{A} U^{\dagger}$. Since $U$ is a Clifford unitary operator, $P_{A}(t)$ will be also a Pauli operator. Hence, we may write it as follows:
\begin{align}
P_{A}(t) \simeq \Lambda_{C} (P_{A}) \otimes \Lambda_{D} (P_{A})  \qquad P_{A} \in \mathrm{Pauli}_{A}
\end{align}
where $\Lambda_{C} (P_{A})$ and $\Lambda_{D} (P_{A}) $ are Pauli operators supported on $C,D$ respectively. Here we used ``$\simeq$'' to denote an equality up to a $U(1)$ phase factor. One can interpret the resulting Pauli operator $\Lambda_{D} (P_{A})$ on $D$ as an output from a map from Pauli operators on $A$ to those on $D$:
\begin{align}
\Lambda_{D} : \mathrm{Pauli}_{A} \rightarrow \mathrm{Pauli}_{D}.
\end{align}
We will call $\Lambda_{D}$ a \emph{forward time-evolution map}. 

Let us examine some basic properties of the forward time-evolution map $\Lambda_{D}$. One can easily verify the following statement. 

\begin{lemma}\label{lemma:linear}
The forward time-evolution map $\Lambda_{D}$ is linear. Namely, we have
\begin{align}
\Lambda_{D} ( P_{A} P_{A}' )\simeq \Lambda_{D}(P_{A})\Lambda_{D}(P_{A}').
\end{align}
Also, the dagger operation is preserved: 
\begin{align}
\Lambda_{D} ( P_{A}^{\dagger} ) \simeq \Lambda_{D} ( P_{A} )^{\dagger}.
\end{align}
\end{lemma}


\subsection{Recoverability}

Next, let us study the recoverability. Recall that, if perfect decoupling occurs, then we will have $\rho_{RC}=\rho_{R}\otimes \rho_{C}$. Hence, by computing the entanglement entropy of $\rho_{RC}$, one can assess the decoupling property (lemma~\ref{lemma:criteria}). Here, it is convenient to compute the R\'{e}nyi-$2$ entropy, by recalling that, for a Clifford dynamics, the R\'{e}nyi-$2$ entropy and the Von Neumann entropy match. We have
\begin{align}
\Tr (\rho_{RC}^2) =\ \figbox{1.2}{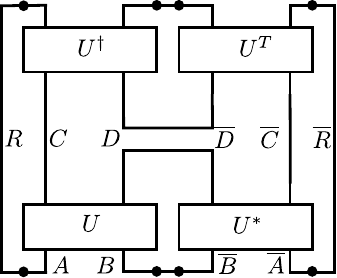} \  = \ \frac{1}{d_{A}^2} \sum_{P_{A}} \ \figbox{1.2}{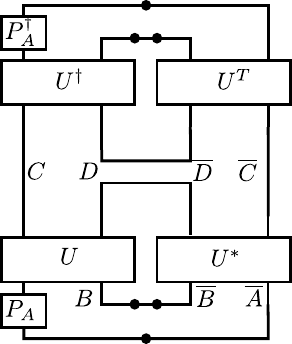} 
\ = \ \frac{1}{d_{R}d_{C}} \sum_{P_{A}} \figbox{1.2}{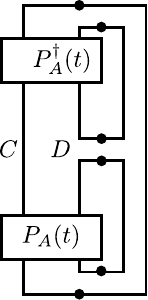} \notag
\end{align}
where black dots represent factors of $\frac{1}{\sqrt{d_{\text{dim}}}}$ for normalizations. Here we inserted the summation over all the Pauli operators on $A$ by using 
\begin{align}
\frac{1}{d_{A}}\sum_{P_{A}} P_{A}\otimes P_{\overline{A}}^{\dagger}=\text{SWAP}_{A\overline{A}}. 
\end{align}

We see that a non-trivial contribution to $\Tr (\rho_{RC}^2)$ comes from a Pauli operator $P_{A}$ such that $\Lambda_{D}(P_{A})=I_{D}$. Hence we arrive at the following lemma:

\begin{lemma}
The entanglement entropy of $\rho_{RC}$ is given by
\begin{align}
S^{(1)}_{RC} =S^{(2)}_{RC} = - \log \Tr (\rho_{RC}^2) =  \log \frac{d_C d_R}{\mathcal{N}_{I_{D}}}
\end{align}
where $\mathcal{N}_{I_{D}}$ is defined by
\begin{align}
\mathcal{N}_{I_{D}} \equiv \text{number of $P_{A}\in \mathrm{Pauli}_{A}$ such that ``$\Lambda_{D}(P_{A})\simeq I_{D}$''}.
\end{align}
\end{lemma}

So, perfect decoupling (and perfect recovery) can be achieved if and only if $\mathcal{N}_{I_{D}}=1$. Finally, one can show that 
\begin{align}
\mathcal{N}_{I_{D}} = 1 \quad \Leftrightarrow \quad \text{$\Lambda_{D}$ is a one-to-one map} \label{eq:statement}
\end{align}
by using lemma~\ref{lemma:linear}. 
Namely, note $\mathcal{N}_{I_{D}}>1$ implies that $\Lambda_{D}$ is not one-to-one. Also, if $\Lambda_{D}$ is not one-to-one, then we have different Pauli operators $P_{A}$ and $P_{A}'$ such that $\Lambda_{D}(P_{A})\simeq \Lambda_{D}(P_{A}')$, which suggests $\Lambda_{D}(P_{A} P_{A}'^{\dagger}) \simeq I_{D}$ and $\mathcal{N}_{I_{D}}>1$. This proves the statement in Eq.~\eqref{eq:statement}. 

The central result of this section is summarized below.

\begin{theorem}
The Hayden-Preskill recovery problem with a Clifford dynamics admits a perfect recovery if and only if $\Lambda_{D}$ is a one-to-one map. 
\end{theorem}

Let us discuss some physical intuition of this result. If a perfect decoupling occurs, then, by looking at Pauli operators $\Lambda_{D}(P_{A})$ supported on $D$, one can deduce the original Pauli operator $P_{A}$ on $A$. As such, the recoverability of quantum information is equivalent to recoverability of the original local operators on $A$ from the time-evolved operators accessed from $D$. Here it is worth recalling that a classical version of the Hayden-Preskill thought experiment can be understood in a similar manner. Namely, recovery is possible when each binary input on $A$ generates unique binary output on $D$~\cite{Hayden07}.

Finally, let us recall that, if $U$ is a random Clifford unitary, a perfect decoupling, $\rho_{RC}=\rho_{R}\otimes \rho_C$, occurs with probability $1-O(\frac{d_{A}^2}{d_{D}^2})$. This is due to the fact that the map $\Lambda_{D}$ is likely to generate non-overlapping Pauli operators $\Lambda_{D}(P_{A})$ when $d_{D}\gg d_{A}$.

\section{Recovery algorithm}\label{sec:recovery}

In this section, we will present a recovery algorithm for the Clifford Hayden-Preskill problem. The basic strategy is to perform generalized Bell measurement and apply appropriate feedback operations based on measurement results. Here, we compute the measurement probability and present a method of constructing feedback operations by looking at the operator growth, namely the forward time-evolution map $\Lambda_{D}(P_{A})$.

\subsection{Measurement probability}

Recall that projecting $D\overline{D}$ onto $|\text{EPR}\rangle_{D\overline{D}}$ will achieve a probabilistic recovery~\cite{Yoshida:2017aa}. To promote this strategy to a deterministic one, we will perform \emph{generalized Bell measurements} with the following basis states:
\begin{align}
|Q_{D}\rangle \equiv (Q_{D}\otimes I_{\overline{D}}) |\text{EPR}\rangle_{D\overline{D}} \qquad Q_{D} : \text{Pauli operator on $D$}.
\end{align}
Note that there are $d_{D}^2$ different Pauli operators and $|Q_{D}\rangle $ form a complete set of mutually orthogonal states on $D\overline{D}$. The idea is to perform appropriate feedback operation based on the outcome of the generalized Bell measurement. 

The probability of measuring $|Q_{D}\rangle$ can be explicitly evaluated as follows:
\begin{align}
\mathrm{Prob}(|Q_{D}\rangle)= \ \figbox{1.2}{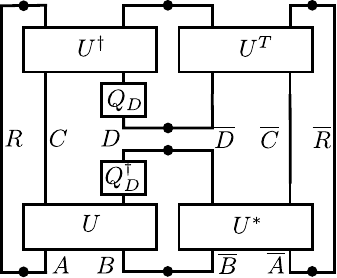} \ = \ \frac{1}{d_{A}^2}\sum_{P_{A}} \figbox{1.2}{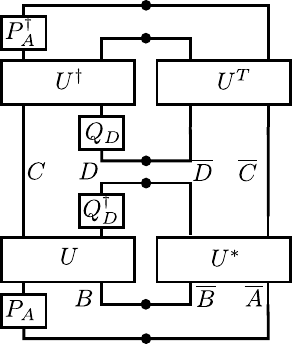} \ = \ \frac{1}{d_{A}^2}\sum_{P_{A}} \figbox{1.2}{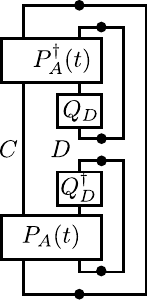} \notag
\end{align}
where we again inserted the summation over Pauli operators on $A$. 

We see that non-trivial contribution to $\mathrm{Prob}(|Q_{D}\rangle)$ comes only when $P_{A}(t)$ has $Q_{D}$ on $D$. Hence, we arrive at the following lemma.

\begin{lemma}
In the Clifford Hayden-Preskill problem, the probability of measuring $|Q_{D}\rangle$ on $D\overline{D}$ is  
\begin{equation}
\mathrm{Prob}(|Q_{D}\rangle)= \frac{1}{d_{A}^2}\mathcal{N}_{Q_{D}}
\end{equation}
where $\mathcal{N}_{Q_{D}}$ is defined by
\begin{align}
\mathcal{N}_{Q_{D}} \equiv \text{number of $P_{A}$ such that ``$\Lambda_{D}(P_{A})=Q_{D}$''}.
\end{align}
\end{lemma}

So, if $\Lambda_{D}(P_{A})\not= Q_{D}$ for all $P_{A}$, then $|Q_{D}\rangle$ will never be measured. The probability of measuring $|\text{EPR}\rangle_{D\overline{D}}$ is $\frac{\mathcal{N}_{I}}{d_{A}^2}$. Hence, a projection onto $|\text{EPR}\rangle_{D\overline{D}}$ will achieve a perfect recovery (which would happen when $\mathcal{N}_{I}=1$) only with probability $\frac{1}{d_{A}^2}$ as pointed out in~\cite{Yoshida:2017aa}.

\subsection{Feedback operation}

Next, let us construct the appropriate feedback operation.  Suppose that a perfect decoupling occurred. There are $d_{A}^2$ possible different measurement outcomes, $|\Lambda_{D}(P_{A})\rangle$, and each will be measured with probability $\frac{1}{d_{A}^2}$. It turns out that, when $|\Lambda_{D}(P_{A})\rangle$ is measured, all we need to do is to apply a feedback Pauli operator $P_{A}$ on $\overline{R}$. That is, depending on the measurement outcome $|\Lambda_{D}(P_{A})\rangle$, we simply need to find $P_{A}$ by looking at the inverse of the map $\Lambda_{D}$. 

To see why this feedback works, note that applying $P_{A}$ on $\overline{R}$ is equivalent to applying $P_{A}^{T}(t)$ on $\overline{D}\overline{C}$. Recalling $P_{A}^{T}(t) = {Q_{C}^{T}}^{(P_{A})} \otimes {Q_{D}^{T}}^{(P_{A})}$, we have
\begin{align}
\figbox{1.2}{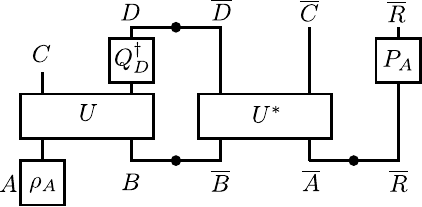} \ = \ 
\figbox{1.2}{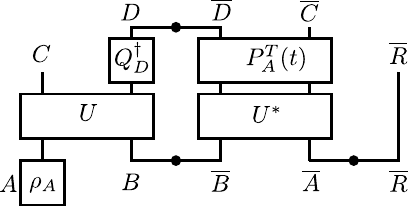} \ = \ \figbox{1.2}{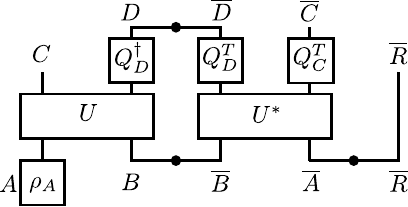}\ . \notag
\end{align}
The actions of $Q_{D}^{\dagger}$ and ${Q_{D}^{T}}$ get cancelled when acting on $|\text{EPR}\rangle_{D\overline{D}}$. Hence the total operation is equivalent to projecting $D\overline{D}$ onto $|\text{EPR}\rangle_{D\overline{D}}$ and apply ${Q_{C}^{T}}^{(P_{A})}$ on $\overline{C}$:
\begin{align}
\figbox{1.2}{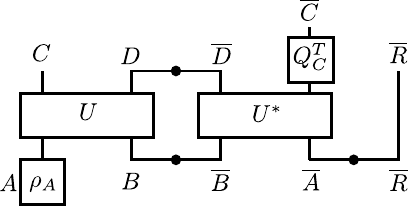}\ .
\end{align}
As such, we achieve a perfect reconstruction fidelity without postselection.

The algorithm is graphically shown below:
\begin{align}
\figbox{1.2}{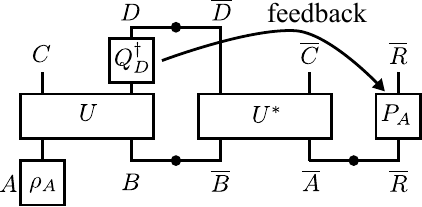} 
\end{align}
and is summarized as follows:

\begin{enumerate}
\item For each Pauli operator $P_{A}$, compute its time-evolution $P_{A}(t) = U P_{A} U^{\dagger}= Q_{C}^{(P_{A})} \otimes Q_{D}^{(P_{A})}$. 
\item Perform a Bell measurement with the basis states $|Q_{D}\rangle$ on $D\overline{D}$. 
\item If $|Q_{D}^{(P_{A})}\rangle$ is measured, then apply $P_{A}$ on $\overline{R}$. The initial state will be reconstructed on $\overline{R}$.
\end{enumerate}

\subsection{Imperfect decoupling}

Finally, let us look at the cases with imperfect decoupling. Namely, we explicitly compute the output state on $R\overline{R}$. Since the output on $R\overline{R}$ does not depend on the measurement result once an appropriate feedback is applied, it suffices to study the case where $|\text{EPR}\rangle_{D\overline{D}}$ was measured. Since the probability of measuring $|\text{EPR}\rangle_{D\overline{D}}$ is $\frac{\mathcal{N}_{I_{D}}}{d_{A}^2}$, the normalized output state is given by
\begin{align}
\rho_{R\overline{R}} =\ \frac{1}{\mathcal{N}_{I_{D}}} \figbox{1.2}{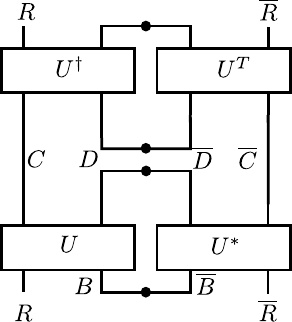}.
\end{align}
Let us expand $\rho_{R\overline{R}}$ by using the following orthonormal complete basis states on $R\overline{R}$:
\begin{align}
|P_{R}\rangle \equiv (P_{R}\otimes I_{\overline{R}}) |\text{EPR}\rangle_{R\overline{R}} \qquad P_{R} \in \mathrm{Pauli}_{R}.
\end{align}
We then have
\begin{align}
\langle P_{R}' | \rho_{R\overline{R}} | P_{R}\rangle = \ \figbox{1.2}{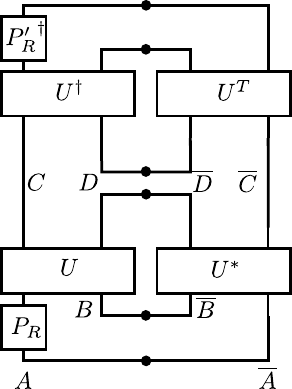} \ .
\end{align}
The nonzero contribution comes only when $P_{R}=P_{R}'$ and $\Lambda_{D}(P_{R})=I_{D}$. So, we arrive at the following result.

\begin{lemma}
The output of the aforementioned recovery algorithm for the Clifford Hayden-Preskill problem is
\begin{align}
\rho_{R\overline{R}} = \frac{1}{\mathcal{N}_{I_{D}}} \sum_{P_{R}:\Lambda_{D}(P_{R})=I_{D}} |P_{R}\rangle\langle P_{R}|.
\end{align}
\end{lemma}

One can check that, for the perfect decoupling case $\mathcal{N}_{I_{D}}=1$, we recover $\rho_{R\overline{R}} =|\text{EPR}\rangle\langle \text{EPR}|_{R\overline{R}}$.

\section{Logical operators}\label{sec:logical}

In this section, we will construct expressions of logical operators supported on $D\overline{B}$.

\subsection{Backward time-evolution map}

When a perfect decoupling occurs, there must exist expressions of (all the) logical operators supported on $D\overline{B}$. In order to construct logical operators, it is convenient to analyze a backward time-evolution. Given a Pauli operator $Q_{D}$ on $D$, let us consider $Q_{D}(-t)=U^{\dagger}Q_{D}U$, which can be written as follows:
\begin{align}
Q_{D}(-t) \simeq \Omega_{A}(Q_{D}) \otimes \Omega_{B}(Q_{D}) \qquad Q_{D} \in \mathrm{Pauli}_{D}
\end{align}
where $\Omega_{A}(Q_{D})$ and $\Omega_{B}(Q_{D})$ are Pauli operators supported on $A,B$ respectively. One can interpret the resulting Pauli operator $\Omega_{A}(Q_{D})$ on $A$ as an output from a map from Pauli operators on $D$ to those on $A$:
\begin{align}
\Omega_{A}: \mathrm{Pauli}_{D} \rightarrow \mathrm{Pauli}_{A}.
\end{align}
We will call $\Omega_{D}$ a \emph{backward time-evolution map}. One can see that a similar statement as lemma~\ref{lemma:linear} holds for $\Omega_{A}$ as well.

As summarized in the following lemma, logical operators can be constructed from $Q_{D}(-t)$.

\begin{lemma}
Assume that there exists $Q_{D}$ such that $\Omega_{A}(Q_{D})=P_{A}$. Then, a logical operator $\widetilde{P_{A}}$ can be constructed as follows
\begin{align}
\widetilde{P_{A}} = \ e^{-i\theta} \figbox{1.2}{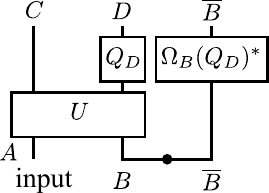} 
\end{align}
where the $U(1)$ phase factor $e^{i\theta}$ results from 
\begin{align}
Q_{D}(-t) = e^{i\theta}\big( P_{A} \otimes \Omega_{B}(Q_{D}) \big).
\end{align}
\end{lemma}

To prove this lemma, observe that applying $Q_{D}$ on $D$ has the same effect as applying $Q_{D}(-t)$ on $A,B$. Hence we have 
\begin{align}
\figbox{1.2}{fig-logical1.pdf} \ = \ e^{i\theta}
\figbox{1.2}{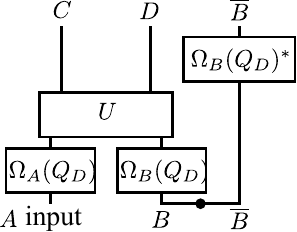} \ = \ e^{i\theta}
\figbox{1.2}{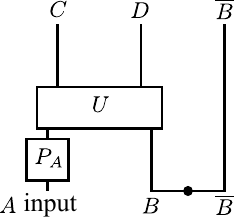}\ 
\end{align}
which has the same effect as applying $e^{i\theta}\Omega_{A}(Q_{D})=e^{i\theta} P_{A}$ on the input state.

\subsection{Existence of $Q_{D}$}

An expression of a logical operator $\widetilde{P_{A}}$ can be obtained if one can find $Q_{D}$ such that $\Omega_{A}(Q_{D})=P_{A}$. The remaining task is to prove that, for a given $P_{A}$, there exists $Q_{D}$ such that $\Omega_{A}(Q_{D})=P_{A}$.

\begin{lemma}\label{lemma:existence}
In the Clifford Hayden-Preskill problem with perfect decoupling, for a given Pauli operator $P_{A}\in \mathrm{Pauli}_{A}$, there always exists $Q_{D}\in \mathrm{Pauli}_{D}$ such that 
\begin{align}
\Omega_{A}(Q_{D})=P_{A}. 
\end{align}
\end{lemma}

To prove this lemma, it is convenient to introduce the following matrix which records the values of OTOC functions:
\begin{align}
\alpha_{P_{A},Q_{D}} \equiv \langle P_{A}(t) Q_{D}(0) P_{A}^{\dagger}(t) Q_{D}^{\dagger}(0) \rangle = \frac{1}{d}\Tr \Big( P_{A}(t) Q_{D}(0) P_{A}^{\dagger}(t) Q_{D}^{\dagger}(0)  \Big)
\end{align}
where $P_{A}\in \mathrm{Pauli}_{A}$ and $Q_{D}\in \mathrm{Pauli}_{D}$. One can rewrite $\alpha_{P_{A},Q_{D}}$ as 
\begin{align}
\alpha_{P_{A},Q_{D}}  =  \langle P_{A}(0) Q_{D}(-t) P_{A}^{\dagger}(0) Q_{D}^{\dagger}(-t) \rangle .
\end{align}
For a Clifford dynamics $U$, we always have $|\alpha_{P_{A},Q_{D}}| = 1$.

Given the values of $\alpha_{P_{A},Q_{D}}$ for a fixed unknown $P_{A}$, one can deduce $P_{A}(t)$. To see this, let us introduce the following matrix:
\begin{align}
F_{P,Q} \equiv  \langle PQP^{\dagger}Q^{\dagger}\rangle
\end{align}
where $P,Q$ are Pauli operators. In the appendix, we prove that this matrix is invertible.

\begin{lemma}
We have 
\begin{align}
\frac{1}{d^2}\sum_{Q\in \mathrm{Pauli}}  F_{P,Q} F_{Q,R} = \delta_{P,R},
\end{align}
or equivalently
\begin{align}
\frac{1}{d^2}\sum_{Q\in \mathrm{Pauli}} \langle PQ P^{\dagger} Q^{\dagger}\rangle\langle QR  Q^{\dagger} R^{\dagger} \rangle = \delta_{P,R}.
\end{align}
\end{lemma}

The matrix $F_{P,Q}$ plays a role similar to the Fourier transformation for Pauli operators. By utilizing this relation, we find
\begin{align}
\frac{1}{d_{D}^2} \sum_{Q_{D}\in \mathrm{Pauli}_{D}} \alpha_{P_{A},Q_{D}} F_{Q_{D}, R_{D}} = \delta_{\Lambda_{D}(P_{A}),Q_{D}}
\end{align}
and
\begin{align}
\frac{1}{d_{A}^2} \sum_{P_{A}\in\mathrm{Pauli}_{A}} F_{R_{A},P_{A}} \alpha_{P_{A},Q_{D}} = \delta_{\Omega_{A}(Q_{D}), R_{A}}. \label{eq:logical_solve}
\end{align}

Finally, let us prove lemma~\ref{lemma:existence}. Suppose that there exists $R_{A}\in \mathrm{Pauli}_{A}$ such that $\delta_{\Omega_{A}(Q_{D}), R_{A}}=0$ for all $Q_{D}\in \mathrm{Pauli}_{D}$. Then we have 
\begin{align}
\sum_{P_{A}\in \mathrm{Pauli}_{A}} F_{R_{A}, P_{A}} \alpha_{P_{A},Q_{D}} = 0 \qquad \text{for all $Q_{D}\in \mathrm{Pauli}_{D}$}.
\end{align}
Multiplying $F_{Q_{D},S_{D}}$ and summing over $Q_{D}$, we obtain 
\begin{align}
\sum_{P_{A}\in \mathrm{Pauli}_{A}}\sum_{Q_{D}\in \mathrm{Pauli}_{D}} F_{R_{A}, P_{A}} \alpha_{P_{A},Q_{D}} F_{Q_{D},S_{D}} =
 0 \qquad \text{for all $S_{D}\in \mathrm{Pauli}_{D}$}.
\end{align}
Hence we have
\begin{align}
\sum_{P_{A}\in \mathrm{Pauli}_{A}} F_{R_{A}, P_{A}} \delta_{\Lambda_{D}(P_{A}), S_D} =  0 \qquad \text{for all $S_{D}\in \mathrm{Pauli}_{D}$}.
\end{align}
For a perfect decoupling case, for a given $S_{D}$, there must exists one and only one $P_{A}'$ such that $\Lambda_{D}(P_{A}')=S_{D}$. So we have 
\begin{align}
F_{R_{A}, P_{A}'}  =  0.
\end{align} 
However, this contradicts with the fact that $|F_{R_{A}, P_{A}'}|=1$. This completes the proof.

Finally, it is worth noting that a Pauli operator $Q_{D}$ such that $\Omega_{A}(Q_{D})=P_{A}$ can be found by solving Eq.~\eqref{eq:logical_solve}, which can be performed efficiently on a classical computer.



\section{Recovery by local measurement}\label{sec:local}

In the previous section, we showed that a deterministic recovery is possible for a Clifford dynamics by performing generalized Bell measurements with entangled basis states. 
In this section, we show that the generalized Bell measurement can be replaced with local (spatially separated) Pauli measurements on $D$ and $\overline{D}$ as long as $d_{D}\geq d_{A}^2$ (instead of $d_{D}\geq d_{A}$). 
We will also briefly discuss its potential application as a possible form of quantum many-body teleportations.

\subsection{Recovery algorithm}

The basic strategy is to measure $D$ and $\overline{D}$ separately with Pauli basis states and apply some appropriate feedback based on the measurement results. Let us clarify our setup and introduce some notations. Without loss of generality, we may assume that both $D$ and $\overline{D}$ are measured in Pauli $Z$-basis states. Also, let us assume that $D, \overline{D}$ have $n_{D}$ qubits respectively and label them by $D_{j}, \overline{D}_{j}$ with $j=1,\cdots, n_{D}$. We denote the measurement results for  $D_{j},\overline{D_{j}}$ by $m_{j}, \overline{m_{j}}=0,1$ respectively, corresponding to $Z=1, -1$. Here we use $\bf{m},\bf{\overline{m}}$ to denote $m_{j}, \overline{m_{j}}$ collectively. It will be convenient to introduce the sum of the measurement outcomes:
\begin{align}
s_{j} \equiv m_{j} + \overline{m_{j}} \qquad \bf{s}\equiv\bf{m}+\bf{\overline{m}} \qquad (\text{mod $2$}).
\end{align}

Again, it is useful to study the operator growth. Consider the time-evolved operator $P_{A}(t)=Q_{C}^{(P_{A})}\otimes Q_{D}^{(P_{A})}$. Here we are particularly interested in whether $Q_{D}^{(P_{A})}$ commutes with Pauli $Z$ operators on $D$ or not. Namely, one can assign a binary string $\textbf{s}^{(P_{A})}$ such that 
\begin{equation}
\begin{split}
&s^{(P_{A})}_{j} = 0 \qquad \text{for}\quad Q_{D_{j}}^{(P_{A})}=I,Z\\
&s^{(P_{A})}_{j} = 1 \qquad \text{for}\quad Q_{D_{j}}^{(P_{A})}=X,Y.
\end{split}
\end{equation}
In other words, we assign $s^{(P_{A})}_{j} =0$ ($s^{(P_{A})}_{j} =1$) when $Q_{D}^{(P_{A})}$ commutes (anti-commutes) with $Z_{j}$. From these commutation relations, one can construct the following linear map $\Lambda_{Z}$:
\begin{align}
\Lambda_{Z}(P_{A}) \equiv \textbf{s}^{(P_{A})}.
\end{align}
One can verify that the map $\Lambda_{Z}$ is linear
\begin{align}
\Lambda_{Z} ( P_{A} P_{A}' ) = \textbf{s}^{(P_{A})} + \textbf{s}^{(P_{A}')} \qquad \text{(mod $2$)}.
\end{align}

The probability of measuring $\bf{m},\bf{\overline{m}}$ is given by
\begin{align}
\figbox{1.2}{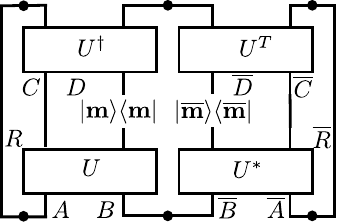} \ = \ \frac{1}{d_{A}^2}\sum_{P_{A}}\figbox{1.2}{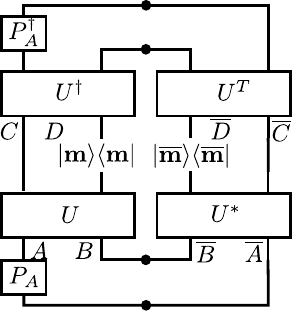} \ = \ \frac{1}{d_{A}^2}\sum_{P_{A}}\figbox{1.2}{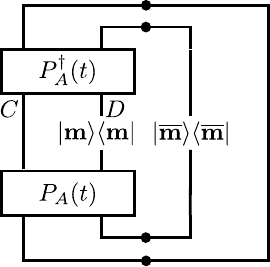}\ .
\end{align}
Let us evaluate the summand for each $P_{A}$. Consider the time-evolved operator $P_{A}(t)=Q_{C}^{(P_{A})}\otimes Q_{D}^{(P_{A})}$. We can see that non-zero contribution (which is $\frac{1}{d_{D}}$) arises only when
\begin{equation}
\begin{split}
& Q_{D_{j}}^{(P_{A})} = I_{j}, Z_{j} \qquad\ \text{for}\quad m_{j}= \overline{m_{j}}\quad (s_{j}=0) \\
&Q_{D_{j}}^{(P_{A})} = X_{j}, Y_{j} \qquad \text{for} \quad m_{j}\not= \overline{m_{j}}\quad (s_{j}=1) 
\end{split}
\end{equation}
since $I_{j},Z_{j}$ acts trivially (up to a phase) on $|\bf{m}\rangle$ and $|\bf{\overline{m}}\rangle$. Hence, we see that 
\begin{equation}
\text{$|\bf{m}\rangle, |\bf{\overline{m}}\rangle$ are measured with probability $\frac{1}{d_{D}}\frac{1}{d_{A}^2}\mathcal{N}_{\bf{s}} $}
\end{equation}
where $\mathcal{N}_{\bf{s}}$ is defined by
\begin{align}
\mathcal{N}_{\bf{s} } \equiv \text{number of $P_{A}$ such that ``$\Lambda_{Z}(P_{A})=\bf{s}$''}.
\end{align}

Having computed the measurement probabilities, let us compute the recovery fidelity. Recall that, for a generic scrambling unitary, one can achieve a nearly perfect recovery by post-selecting the measurement results to be $\bf{m}=\bf{\overline{m}}$ ($\bf{s}=0$). Let us verify a similar statement for a Clifford dynamics. When $\bf{m}$ and $\bf{\overline{m}}$ are measured, the normalized wavefunction after the measurement is given by (as long as $\mathcal{N}_{\bf{s}}\not=0$)
\begin{align}
\sqrt{\frac{d_{D}d_{A}^2}{\mathcal{N}_{\bf{s}}} }\ \figbox{1.2}{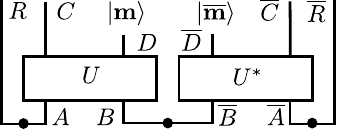}\ .
\end{align}
The EPR fidelity of $R,\overline{R}$ is given by 
\begin{align} 
\frac{d_{D}d_{A}^2}{\mathcal{N}_{\bf{s}}} \ \figbox{1.2}{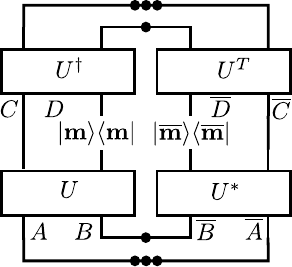} \ = \ \frac{1}{\mathcal{N}_{\bf{s}}} \delta_{\bf{s},\bf{0}}.
\end{align}
Hence, by post-selecting the measurement result to be $\bf{m}=\bf{\overline{m}}$, a perfect recovery is possible as long as $\mathcal{N}_{\bf{0}}=1$. In other words, the map $\Lambda_{Z}$ must be one-to-one. It is worth noting that, when $U$ is a random Clifford with $d_{D}^2 \geq d_{A}$, one can show that the binary strings $\textbf{s}^{(P_{A})}$ are all different with high probability. Hence, a perfect recovery is almost surely possible. 

If the measurement result is $\bf{m}\not=\bf{\overline{m}}$, one needs to apply a certain feedback Pauli operator $P_{A}$. Here $P_{A}$ should be chosen so that $Q_{D_{j}}^{(P_{A})} = X,Y$ at $D_{j}$ with $m_{j}\not= \overline{m_{j}}$. The existence of such $P_{A}$ is guaranteed from the fact that the measurement of $\bf{m},\bf{\overline{m}}$ occurred. Applying $P_{A}$ has an effect of effectively restoring $D,\overline{D}$ to $\bf{m}=\bf{\overline{m}}$. The final outcome of the recovery algorithm can be found as follows 
\begin{align}
\rho_{R\overline{R}} = \frac{1}{\mathcal{N}_{\bf{0}}} \sum_{P_{R}:\Lambda_{Z}(P_{R})=\bf{0}} |P_{R}\rangle\langle P_{R}|.
\end{align}
These results can be proven with minor modifications to proofs in section~\ref{sec:recovery}. Also, it can be shown that to construct relevant logical operators, one needs to consider backward evolution of Pauli-$Z$ operators on $D$. 

The recovery algorithm is graphically shown below:
\begin{align}
\figbox{1.5}{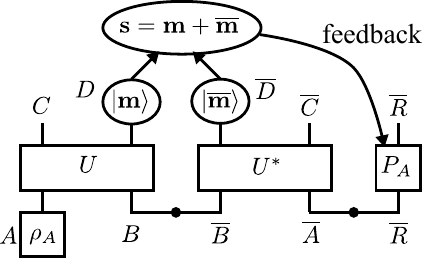}  \label{eq:summary_local}
\end{align}
and is summarized as follows:

\begin{enumerate}
\item For each Pauli operator $P_{A}$, compute its time-evolution $P_{A}(t) = U P_{A} U^{\dagger}= Q_{C}^{(P_{A})} \otimes Q_{D}^{(P_{A})}$, and construct a binary string $\textbf{s}^{(P_{A})}$ which records the commutation with Pauli-$Z$ operators. 
\item Perform $Z$-basis measurements on $D$ and $\overline{D}$ and compute the sum $\bf{s}=\bf{m} + \bf{\overline{m}}$.
\item If $\bf{s}=\textbf{s}^{(P_{A})}$, then apply $P_{A}$ on $\overline{R}$. The initial state will be reconstructed on $\overline{R}$.
\end{enumerate}

A physical interpretation of the linear map $\Lambda_{Z}$ is as follows. When generalized Bell measurements in entangled basis states are performed, one is able to measure the Pauli operator $\Lambda(P_{A})=Q_{D}^{(P_{A})}$ supported on $D$. When local Pauli $Z$-basis measurements are performed, one may only measure whether $Q_{D}$ commutes or anti-commutes with Pauli-$Z$ operator, and hence one obtains less information. 

It is worth noting that this recovery algorithm with local Pauli measurement is useful in studying entanglement structure in random Clifford circuits with local measurements (the so-called monitored circuit) as we shall discuss elsewhere.

\subsection{Many-body quantum teleportation}

The aforementioned recovery algorithm for the Clifford Hayden-Preskill problem with local Pauli basis states provides a novel many-body generalization of quantum teleportation with fault-tolerance. In this protocol, the sender and the receiver performs Pauli measurements on qubits in their own laboratories, and only the classical message of the sender's measurement results need to be communicated to the receiver. Here we briefly discuss its potential advantage over conventional quantum teleportation. 

First, the protocol is robust against possible attempts to eavesdrop the quantum state $|\psi_A\rangle$. Namely, in order to steal $|\psi\rangle_A$, the eavesdropper needs to access a large portion of qubits from the sender's end. When the dynamics $U$ is a random Clifford operator, the number of such qubits $n_{E}$ should be $n_{E}\gtrapprox n-n_{A}$. As such, the Clifford Hayden-Preskill setup may provide a secure quantum communication channel. 

Second, the protocol can detect errors during its implementation. Let us take $D$ to be sufficiently larger than $A$, namely $n_{D}\gg 2n_{A}$. If there is no error, then the measurement sum $\bf{s}$ should coincide with $\bf{s}^{(P_{A})}$ for some Pauli operator $P_{A}$. If this is not the case, then one can deduce that some error has occurred while implementing the protocol. As such, this protocol comes with error detection capability. In fact, we expect that it is possible to not only detect errors but also collect errors by looking at the values of the measurement sum $\bf{s}$. We leave these aspects as future problems.

\section{Outlook}\label{sec:outlook}

Our analysis suggests that measurements with local basis states reveal limited information about the operator growth, namely only about whether the time-evolved operators commute or anti-commute with measured Pauli operators. It is natural to expect that the mutual information $I(R,\overline{B}D)$ is no longer a good criteria when our measurement abilities are restricted. One possible candidate quantity is an algebraic generalization of the mutual information which is based on the relative entropy~\cite{Ohya_Petz_Text} where the algebra may be chosen according to our measurement abilities. It is worth noting that in the appendix of~\cite{Yoshida:2019aa}, it has been observed that OTOC functions can be associated with R\'{e}nyi-$2$ generalization of the relative entropy~\cite{Beigi:2013aa, Muller-Lennert:2013aa, Wilde:2014aa}, which may be useful in further generalizing the Hayden-Preskill problem to an algebraic setups. 

It is useful to note that, while a random Clifford dynamics leads to the Hayden-Preskill recovery phenomena, it does \emph{not} satisfy the rigorous definition of quantum information scrambling as diagnosed by out-of-time order correlation (OTOC) functions. Essentially, this is due to the fact that the absolute value of OTOC functions with respect to Pauli operators remains to be unity whereas, for truly scrambling dynamics, OTOC functions decay to small values for all the local operators. Hence, a random Clifford unitary is an example of non-scrambling systems which still admit the Hayden-Preskill recovery.

In the appendix, it was observed that there is an interesting relation between out-of-time order correlation functions and discrete Wigner functions. There have been a number of interesting works concerning (non-)Cliffordness through resource theoretic arguments, see~\cite{Veitch:2014aa, Delfosse:2015aa, Howard:2014aa, Howard:2017aa} for instance. Hence, it will be interesting to study quantum information scrambling from the viewpoint of resource theory. For this purpose, it will be useful to note that the absolute value of out-of-time order correlation functions, with respect to Pauli operators, remain to be unity when the dynamics is given by Clifford, whereas it decays to small values for non-Clifford dynamics.

A version of the Hayden-Preskill problem for continuous variable systems for Gaussian dynamics has been studied in~\cite{Zhuang:2019aa}, which will be interesting to further study in the context of EAQCCEs. We expect that the decoding algorithms in this note can be applied to continuous variable systems as well. 

\subsection*{Acknowledgment}

Research at the Perimeter Institute is supported by the Government of Canada through Innovation, Science and Economic Development Canada and by the Province of Ontario through the Ministry of Economic Development, Job Creation and Trade.

\appendix

\section{Discrete Wigner function}

In this appendix, we prove that $F_{PR}=\langle PRP^{\dagger}R^{\dagger}\rangle$ is invertible by using some well-known property of discrete Wigner functions. 
Consider a $d$-dimensional Hilbert space $\mathcal{H}=\mathbb{C}^d$. Let us denote generalized Pauli operators on $\mathcal{H}$ by $T_{u}$ with $u=1,\cdots, d^2$. We will prove the following lemma:

\begin{lemma}\label{lemma:inverse}
The matrix $F_{PR}=\langle PRP^{\dagger}R^{\dagger}\rangle$ is invertible, namely
\begin{align}
\frac{1}{d^2}\sum_{Q\in \mathrm{Pauli}} \langle Q P Q^{\dagger} P^{\dagger}\rangle\langle R Q R^{\dagger} Q^{\dagger} \rangle = \delta_{P,R}.
\end{align}
\end{lemma}

To prove this statement, let us consider the following quantum state, which we call the \emph{commutator wavefunction}, defined on a doubled Hilbert space $\mathcal{H}^{\otimes 2}$:
\begin{align}
|\psi_{u}\rangle \equiv \frac{1}{d}\sum_{Q\in \text{Pauli} } \langle T_{u}QT_{u}^{\dagger}Q^{\dagger}\rangle|Q\rangle \qquad
T_{u}\in \text{Pauli}
\end{align}
where $|Q\rangle$ is the Choi state for Pauli operator $Q$:
\begin{align}
|Q\rangle \equiv (Q \otimes I)|\text{EPR}\rangle.
\end{align}
So, the commutator wavefunction $|\psi_{u}\rangle$ records the commutator $\langle T_{u}QT_{u}^{\dagger}Q^{\dagger}\rangle$ as the coefficient of $|Q\rangle$. One can easily verify that $|\psi_{u}\rangle$ is properly normalized. 

The LHS of the equation in lemma~\ref{lemma:inverse} can be obtained by considering an inner product between the commutator wavefunctions $|\psi_{u}\rangle$ and $|\psi_{v}\rangle$:
\begin{align}
\langle \psi_{v} |\psi_{u}\rangle = \frac{1}{d^2} \sum_{Q\in \mathrm{Pauli}} \langle QT_{v}Q^{\dagger}T_{v}^{\dagger}\rangle \langle T_{u}QT_{u}^{\dagger}Q^{\dagger}\rangle.
\end{align}
So, our goal is to show that the commutator wavefunctions are orthonormal; $\langle \psi_{v} |\psi_{u}\rangle =\delta_{v,u}$. 

This can be shown by using a certain property of discrete Wigner functions. Define Wigner operators $W_{u}$ as follows:
\begin{align}
W_{u} \equiv \frac{1}{d}\sum_{Q\in \text{Pauli}} T_{u} Q T_{u}^{\dagger} \qquad T_{u}\in \text{Pauli}
\end{align}
Then we have the following lemma~\cite{Veitch:2014aa}:

\begin{lemma}\label{lemma:Wigner}
The Wigner operators are orthonormal, namely
\begin{align}
\langle W_{u} | W_{v} \rangle = \delta_{u,v}
\end{align}
where $| W_{v} \rangle$ is the Choi representation of $W_{v}$. 
\end{lemma}

The final step is to notice that the aforementioned commutator wavefunction $|\psi_{u}\rangle$ is in fact identical to the Choi representation of $W_{u}$. To see this, observe that 
\begin{align}
|\psi_{u}\rangle  = \frac{1}{d}\sum_{Q\in \text{Pauli}} |T_{u}QT_{u}^{\dagger}\rangle.
\end{align}
This follows from the fact that  $T_{u}QT_{u}^{\dagger}Q^{\dagger}$ is proportional to the identity operator. Hence, we arrive at the following lemma.

\begin{lemma}\label{lemma:relation}
The commutator wavefunction is the Choi representation of the Wigner operator:
\begin{align}
|\psi_{u}\rangle \equiv \frac{1}{d}\sum_{Q\in \mathrm{Pauli} } \langle T_{u}QT_{u}^{\dagger}Q^{\dagger}\rangle|Q\rangle  = |W_{u}\rangle.
\end{align}
\end{lemma}

Finally, with the help of lemma~\ref{lemma:Wigner} and lemma~\ref{lemma:relation}, lemma~\ref{lemma:inverse} can be proven by noting
\begin{align}
\langle \psi_{u} | \psi_{v} \rangle=\langle W_{u} | W_{v} \rangle = \delta_{u,v}.
\end{align}

We have seen that the commutator wavefunction $|\psi_{u}\rangle$, which records the values of commutators $\langle T_{u}QT_{u}^{\dagger}Q^{\dagger}\rangle$ is identical to the Choi representation of the Wigner function $W_{u}$. We expect that, with the help of this identification, various resource theoretic results on (non-)Cliffordness (such as~\cite{Veitch:2014aa}) can be translated into the language of OTOC functions. We leave this as a future problem.

\providecommand{\href}[2]{#2}\begingroup\raggedright\endgroup

\bibliographystyle{utphys}
\bibliography{myref2021.bib}{}
\end{document}